\documentclass[12pt]{article}
\usepackage{graphics}
\newcommand{\lk}{\left( }
\newcommand{\rk}{\right)}

\newcommand{\fft}[2]{{\frac{#1}{#2}}}

\newcommand{\app}{\alpha^\prime}
\newcommand{\del}{\partial}

\newcommand{\bea}{\begin{eqnarray}}
\newcommand{\eea}{\end{eqnarray}}

\usepackage[usenames]{color}

\begin{document}
\begin{titlepage}

\title
{
%\begin{flushright}
%\normalsize {\tt hep-th/yymmnnn}\\
%\end{flushright}
\vfill
\large \bf Drag Force, Jet Quenching, and AdS/QCD
}

\author{
{Eiji Nakano}${}^{1}${}\thanks
{{\tt enakano@ntu.edu.tw}}
,
{Shunsuke Teraguchi}${}^{2}${}\thanks
{{\tt teraguch@phys.sinica.edu.tw}}
and
{Wen-Yu Wen}${}^{1}${}\thanks{{\tt steve.wen@gmail.com}}\\
\normalsize {${}^{1}$ \it Department of Physics and Center for
Theoretical Sciences, National Taiwan University,}\\
\normalsize {\it Taipei 10617, Taiwan}\\
\normalsize {${}^{2}$
\it Institute of Physics, Academia Sinica, Taipei 11529, Taiwan}\\
}
\date{}

\maketitle

\begin{abstract}
In this note, two important transport observables in the RHIC
experiment, damping rate and jet quenching parameter,
are calculated from an AdS/QCD model.  A quark moving in the
viscous medium such as the Quark-Gluon-Plasma is modelled by an
open string whose end point travels on the boundary of a deformed
AdS$_5$ black hole.  The correction introduced via the deformed
AdS$_5$ is believed to help us better understand the data which is
expected to be measured in the RHIC.

\end{abstract}

\vfill

\end{titlepage}

%%%%%%%%%%%%%%%%%%%%%%%%%%%%%%%%%%%%%%%%%%%%%%%%%%%%%%%%%%%%%%%%%%%%%%%%%%%%
\section{Introduction}
Observation of leading-particles with transverse momentum $p_T$ in
the relativistic heavy ion collider (RHIC) provides significant
insights into the matter formed at center of the collision, which
is considered to evolve into quark-gluon plasma (QGP) for
sufficiently high collision energy\cite{RHIC}. In Au+Au
collisions at the center-of-mass energy $\sqrt{s_{NN}}=200$ {GeV}
at Brookhaven it has been observed a strong suppression of
high $p_T$ in detectable hadron spectra relative to that in binary
proton+proton collision, which is quantified by `the nuclear
modification factor' $R_{AA}(p_T)$ \cite{jq1}. This suppression
factor, characterizing in-medium effects, approaches an asymptotic
value at the higher momentum region $p_T \ge 5$ GeV and become
independent of detected particle species.  It indicates that
detected particles with high $p_T$ (hadron jet) have a partonic origin,
i.e., only a quark from a $q\bar{q}$ pair produced near the surface of expanding QGP can
escape to outside and form hadron jets, and another quark recoiled
back to inside of matter suffers from medium-induced energy
(momentum) loss, which accounts for the strong suppression of
back-to-back jet production.

A measurable quantity sensitive to this in-medium energy
loss is so-called jet quenching parameter $\hat{q}$. This quantity
parameterizes a suppression factor $P_f(\hat{q},L,\Delta E)$ which
is the probability of the process that a hard quark radiates
energy of $\Delta E$ to medium during propagation in path $L$.
Since $P_f$ gives main contribution to $R_{AA}$, $\hat{q}$ nicely
reflects the medium energy loss effect at the high-$p_T$
region\cite{jq1}. Also, $\hat{q}$ is determined by the mean square
transverse momentum of quark propagating in unit length (or
effectively in mean-free path $\lambda_f$), $\hat{q}=\langle
\vec{p}_\bot{}^2 \rangle/\lambda_f$. It should be noted that
$\lambda_f$ here characterizes medium opacity, and is expected to
be very small so is shear viscosity in QGP because $\eta_s \propto
\lambda_f$. The ratio of $\eta_s$ to entropy density $s$ extracted
from RHIC date is close to a conjectured minimum bound $\eta_s/s
\ge 1/(4\pi)$ \cite{Kovtun:2004de}, implying that QGP is strongly
coupled. Thus, if one tries to derive $\hat{q}$ from first
principle in quantum field theory, non-perturbative calculation
seems to be required to obtain a reasonable value.

Another observable quantity sensitive to the in-medium energy loss
is a damping rate $\mu$ (or friction coefficient) defined by
Langevin equation, $\dot{p}=-\mu p + f$, subject to a driving force $f$.
$f$ is equivalent to a drag force up to sign provided
$\dot{p}=0$. This quantity also characterizes opacity, or
equivalently energy loss in dissipative processes in medium. One
might think some relation between $\hat{q}$ and $\mu$, both of
which describe the energy loss. By definition, jet quenching
parameter is related to momentum fluctuation $\hat{q}\propto
\langle \vec{p}_\bot{}^2 \rangle$, and drag force is to an
averaged momentum loss escaping to medium $\langle \dot{p} \rangle$.
Therefore, if the medium provides a stochastic force as $f$ to
quark moving therein, these two quantities are related to each
other via the fluctuation-dissipation theorem.

On the other hand, a way to approach this strongly coupled region
of gauge theories is the AdS/CFT correspondence
\cite{Maldacena:1997re}.  AdS/CFT allows us to investigate the
strongly coupled regime of gauge theories where 't Hooft coupling
$\lambda=g_{YM}^2N_c\gg 1$ by using classical gravity in {\sl
effectively} five dimensional AdS background, whose boundary
corresponds to four dimensional gauge theory and one extra radial
dimension should be interpreted as the energy scale of the gauge
theory. In particular, one can consider a finite temperature gauge
theory by introducing a black hole inside the AdS
space\cite{Gubser:1996de,Witten:1998zw}. Using AdS/CFT, there are
earlier attempts to addressing jet quenching parameter $\hat{q}$
\cite{Liu:2006ug,jetq,Chernicoff:2006hi} and damping rate $\mu$
\cite{hkkky,Casalderrey-Solana:2006rq,gubser,drag
force,Chernicoff:2006hi}.  Various solutions of modified AdS
background are discussed, namely, corresponding gauge theory are
modified ${\cal N}=4$ SYM's.  On the other hand, there are several
works \cite{AdS/QCD model,Karch:2006pv,Andreev:2006vy} to
generalize the AdS/CFT correspondence to more realistic QCD. While
there are several works to search elaborate string/supergravity
solutions in full 10 dimensional space-time, whose holographic
dual has properties similar to QCD, there are also many trials to
directly construct five dimensional holographic dual of QCD by
demanding to reproduce desired properties of QCD, which are
referred to the AdS/QCD in our paper. In this paper, we reexamine
the above two transport observables by using one of such AdS/QCD
models. Especially, we adopt the model by Karch et al.
\cite{Karch:2006pv} which possesses a confinement mechanism
with the help of a non-trivial dilaton background.
It is worth mentioning an alternative model proposed by Andreev and Zakharov
\cite{Andreev:2006vy}.
They adopt a deformed AdS$_5$ metric but a  trivial dilaton field.
These two models are shown to be equivalent to each other as long as quadratic
terms like $F^2$ in the effective action are concerned \cite{Andreev:2006vy}.

This paper is organized as follows.  After we introduce the model
being concerned in the next section, we warm up with a simple
calculation of free energy of a static quark in thermal QGP. Then
we carry out the calculation for drag force and jet quenching
parameter in section 4.  In comparison to the calculations based
on the usual AdS/CFT, it is shown that the free energy has {\sl
weaker} dependence on temperature, and we observe {\sl
bigger} damping rate and {\sl smaller} jet quenching
parameter.
\footnote{ The statement here seems to be conflicting.
This is because we are comparing our result with the calculation
based on different values of t'Hooft coupling $\lambda$,
where $\lambda=10$ was used for damping rate calculation and
$\lambda=6\pi$ for jet quenching parameter in their
analysis. }

\section{Model}
The action of this model\cite{Karch:2006pv} is given by the
following form,
\bea
S=\int d^5x\sqrt{-G}e^{-\Phi}{\cal L},
\eea
where ${\cal L}$ is a five dimensional Lagrangian density (whose
detail is irrelevant in our calculation) and $\Phi$ is the dilaton
field and is assumed to be proportional to the square of fifth
coordinate $z$. The five dimensional metric $G_{MN}$ is given by
the usual $AdS_5$ metric. It has been argued that this model
reproduces Regge behavior $m_{n,S}^2\sim \sigma_{\rm QCD} (n+S)$.
Where the QCD string tension $\sigma_{\rm QCD}\simeq 0.93{\rm
GeV}^2$ is properly adjusted by choosing a suitable value for the
coefficient of $z^2$ in the dilaton field profile. In the later
discussion, our calculation is based on the following
correspondence between Wilson loop $W(C)$ in the gauge theory and
on-shell worldsheet action $\cal S$,
\bea
\left<W(C)\right>\sim
e^{-\cal S}. \eea where the Nambu-Goto action ${\cal S}$ is given by
\bea \label{NGaction}
{\cal S}=\frac{1}{2\pi\app}\int{d\tau d\sigma
e^{\Phi/2}\sqrt{-\det \del_aX^M \del_bX^N G_{MN}}}.
\eea
Here we
have assumed that the metric of this model is given in Einstein frame,
thus we have the extra dilaton contribution
in front of the usual Nambu-Goto action.

In order to use the above formula,
we must fix some parameters
which are irrelevant in the discussion of the paper \cite{Karch:2006pv}.
The AdS metric and the dilaton field are given by
\footnote{
One may consider a different AdS/QCD model
as \cite{Andreev:2006vy} based on the following modified AdS metric,
\bea
ds^2&=&h(z)\frac{R^2}{z^2}(-dt^2+d\vec{x}^2+dz^2),
\eea
with some suitable function $h(z)$.
However, in our calculation, $h(z)$ will appear only in the combination of
$h(z)e^{\Phi(z)/2}$ as one can easily seen from equation (\ref{NGaction}).
Though we only consider the case of $h(z)=1$ in the text,
we can easily recover the results for non trivial $h(z)$.}
\bea
ds^2&=&\frac{R^2}{z^2}(-dt^2+d\vec{x}^2+dz^2),\\
\Phi(z)&=&cz^2, \label{ansatz_dam}\eea where the coefficient
$c$ is estimated as $c=\sigma_{\rm QCD}/4\simeq 0.23 {\rm GeV}^2$
in the present model. $R$ is the {\sl radius} of AdS
space-time. One can fix the ratio $R^2/\app$ by matching the
linear potential from Wilson loop with the QCD string tension
$\sigma_{\rm QCD}$. Such an estimation was performed in
\cite{Andreev:2006vy} for another proposal of AdS/QCD. Here for
the model in concern, we have 
\bea \frac{R^2}{\app} =\frac{4
\pi}{{\rm e}\, a^2 c}\simeq 3.63, 
\label{fitting1}\eea 
where $a$ is a
parameter in the Cornell potential for quark-antiquark
relative separation $r$,
$V(r)=-\frac{\kappa}{r}+\frac{1}{a^2}r+C_0$ \cite{CorPot}. In the
context of the original AdS/CFT, this ratio corresponds to square
root of t'Hooft coupling $\lambda$. As a result, this value
$\lambda\sim(\frac{R^2}{\app})^2\simeq 13.2$ is close to those in
the earlier calculations of transport observables where
$\lambda\sim 10-18$. Notice that we have fixed this value from the
IR behavior in the confinement phase, instead of tuning it
by hand to the value in the energy scale at issue.

\begin{figure}
\center{
\includegraphics{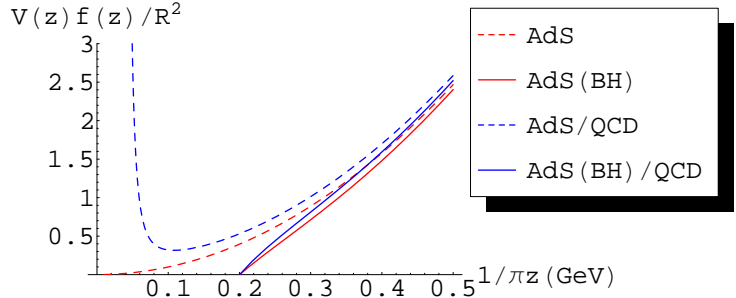}}
\caption{Plot of red shift v.s. energy scale $(1/\pi z)$. At
a temperature of $200$ MeV introduced by black hole (BH), we
plot ${\cal V}(z)f(z)$ for both AdS/CFT and AdS/QCD models. For
comparison, we also plot those at zero temperature. Notice their
differences in the low energy scale.} \label{fig1}
\end{figure}

In this paper, we shall investigate this model with finite
temperature. In order to introduce temperature in the model, a
natural guess is to incorporate black hole into the metric in
(\ref{ansatz_dam}) whose Hawking temperature corresponds to the
temperature at issue:
\begin{eqnarray}\label{ansatz}
&&ds^2=\frac{R^2}{z^2}(-f(z)dt^2+d\vec{x}^2+f(z)^{-1}dz^2),\\
&&f(z)=1-\frac{z^4}{z_T^4}.
\end{eqnarray}
The function $f(z)$ indicates the existence of a black hole with
its event horizon at $z=z_T$. As in the dictionary of AdS/CFT
correspondence, the Hawking temperature is given by the
size of horizon for large AdS black hole,
\begin{equation}
T=\frac{1}{\pi z_T}.
\end{equation}

The property of strings in this geometry can be intuitively grasped
by focusing on the effective redshift factor,
\begin{equation}
{\cal V}(z)f(z)\equiv
|e^{\Phi(z)/2}G_{tt}|=e^{cz^2}\frac{R^2}{z^2}f(z).
\end{equation}
In Fig.~\ref{fig1}, we plot this effective redshift factor for several geometries.
Note that, in any case, the geometries approach the usual AdS$_5$ at the boundary
and its holographic dual gauge theory behaves like a conformal theory in
UV region. On the other hand, the large effective redshift in the IR region of AdS/QCD model
without black hole indicates the confinement of quarks.

\section{Static quark solutions}

\begin{figure}
\center{
\includegraphics{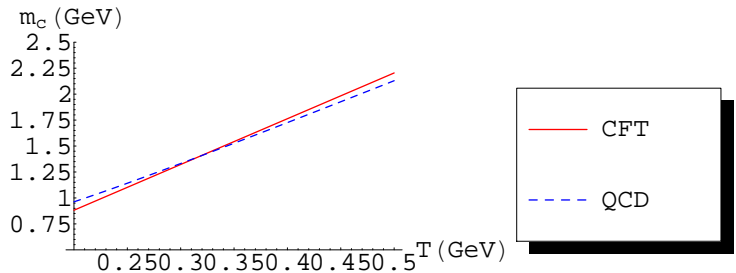}}
\caption{Regulated free energy of charm quark is identified as
quark mass $m_c$ in thermal background. Position of flavor brane
(relative to the horizon) in each model is adjusted to match
experimental data $m_c\simeq 1.4$GeV at $T=318$MeV. $\lambda=10$
is chosen for the CFT model.} \label{fig2}
\end{figure}

Here we consider a static configuration of heavy quarks. Simply by
argument based on symmetry, we learn that while one end of string
is attached to the quark, the other end straightly reaches to the
horizon along $z$-direction. This static configuration gives rise
to the expectation value of a Wilson line in thermal ${\cal N}=4$
super Yang-Mill theory and the free energy is given by the total
string mass \footnote{To be more precise, we consider a
rectangular Wilson loop ${\cal C}$ formed by propagation of a pair
of static heavy quarks along time $T$.  In the limit $T\to
\infty$, the expectation value of Wilson line is given by
$<W({\cal C})>\sim e^{-S}=e^{-TF_{q}}$, where $F_q$ is the
free energy between this pair of quarks and could be understood as
total mass of string connected in between quarks.  While quarks
are widely separated, with the appearance of black hole, a single
heavy quark could be isolated due to a constant potential
(deconfining phase) and $F_q$, upon regularization, could be seen
as quark mass at finite temperature.}. This quantity, however, is
divergent due to the infinite red shift close to the AdS$_5$
boundary.  One way to render it finite is to locate quarks on a
separate {\sl flavor} brane at finite distance $z=z_m$, serving as
a UV regulator\cite{hkkky}. In this way, one obtains the
free energy of static quark $F_q$ as a function of temperature at
a given location of corresponding flavor brane $z_m$,
\begin{equation}
F_{q}(T) =\frac{1}{2\pi\alpha'}
\int_{z_m}^{z_T}{e^{\Phi/2}\sqrt{-G_{tt}G_{zz}}}dz.
\end{equation}
In Fig.~\ref{fig2}, it is shown that temperature dependence of
regulated free energy of charm quark in the present AdS/QCD
model  is {\sl weaker} than that in the usual AdS/CFT.

\section{Moving quark solutions}
\subsection{Drag force calculation} 
\begin{figure}
\center{
\includegraphics{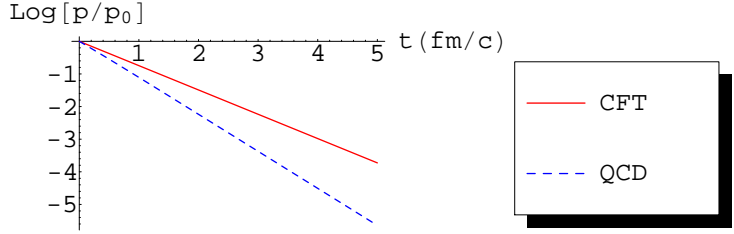}}
\caption{Momentum $p$ v.s. time $t$.  Simulation is done
with charm quark mass $m_c=1.35$ GeV/c at temperature $200$ MeV,
where the initial quark momentum is estimated to be $p_0=10$ GeV/c.
The dashed curve is our result for AdS/QCD and the solid one is for usual
AdS/CFT with $\lambda=10$.} \label{dg1}
\end{figure}

Here we consider a heavy quark
moving on the boundary QCD but with a string tail into the AdS
bulk.  The dissipation of this quark in the QGP is described by the drag force,
which is conjectured to be associated with
a string tail in the fifth dimension.
Now we consider a quark moving along constant $x^2$-$x^3$ plane
and subject to a driving force $f_1$ such that
\begin{equation}
\fft{dp_1}{dt}=-\mu p_1 +f_1.
\end{equation}
One can read the drag force directly from
$f_1$ for constant speed trajectory,
whose corresponding string configuration in the effective five dimensions is given by
\begin{equation}
x^1=vt+\xi{(z)}.
\end{equation}
Here, the coordinates of string worldsheet
$(\tau,\sigma)$ are chosen to be the spacetime coordinates $(t,z)$.
The string action is given by
\begin{equation}
S=\frac{1}{2\pi\alpha'}\int{d\tau d\sigma
{\cal V}[z]\sqrt{1+f{\xi^{'}}^2-f^{-1}v^2}}.
\end{equation}
With the help of the existence of
a conserved energy-momentum current $\Pi$,
one obtains the configuration of string tail as
\begin{equation}
\xi^{'}=\pm\fft{\Pi}{f}\sqrt{\frac{f-v^2}{{\cal V}^2f-\Pi^2}}.
\end{equation}
Given the boundary condition at $z=0$ and $z\to z_T$, one is able
to solve for $\Pi$ and thus integrate to gain $\xi$. Equivalently
one can also impose the reality condition on $\xi^{'}$ and find
out that
\begin{equation}
\Pi={\cal V}(z_v) v, \qquad z_v^4=(1-v^2)z_T^4
\end{equation}
The drag force is then given by \cite{gubser}
\begin{equation}
-f_1=-\fft{1}{2\pi\alpha'}\Pi=-\fft{\pi
T^2R^2}{2\app}e^{\fft12\Phi(z_v)}\fft{v}{\sqrt{1-v^2}},
\end{equation}
The first equality can be understood in the following way: the
energy-momentum flow $\Pi$ along spatial worldsheet direction is
exactly the momentum change of moving quark, hence the applying
force. In the opaque environment such as the QGP, this
momentum change is due to drag force.  The work done by drag force
could be better pictured as the energy-momentum flowing along the
string tail, and finally dumped into the black hole with no
return. The minus sign in front of $f_1$ refers to the
convention for the drag force against the motion of quark. After
applying the AdS/QCD model \cite{Karch:2006pv}, the damping
rate now becomes velocity-dependent, which has the form at each
instantaneous moment
\begin{equation}
\mu(v,T)=\frac{\pi T^2
R^2}{2m\app}e^{\frac{c\sqrt{1-v^2}}{2\pi^2T^2}},
\end{equation}
where $m$ is the mass of quark. This expression reduces to the one
for AdS/CFT case with $c\rightarrow 0$ and $R^2/\app\rightarrow
\sqrt{\lambda}$ \cite{hkkky,gubser}. In Fig.~\ref{dg1}, we plot
the time evolution of momentum for charm quark at temperature
$200$ MeV, where the initial quark momentum is estimated to
be $10$GeV/c. We see that QGP is {\sl more viscous} in this model
than that shown in \cite{gubser}.

\subsection{Jet quenching parameter calculation}
Here we consider a null-like rectangular Wilson loop $C$ formed by
a pair of quark-antiquark with separation $L$ travelling along
light-cone time duration $L^-$.  The jet quenching parameter
$\hat{q}$ is related to the Wilson loop expectation value by
\cite{Liu:2006ug,Liu:2006he}
\begin{equation}
\langle{\cal W}^A(C)\rangle\simeq
\exp\left(-\frac{1}{4\sqrt{2}}\hat{q}L^-L^2\right), \label{Adsco1}
\end{equation}
where the superscript $A$ denotes the adjoint representation.
Using AdS/CFT correspondence, one is able to calculate it
in the fundamental representation as
\begin{equation}
\langle{\cal W}^F(C)\rangle \simeq  e^{-S_I},
\label{Adsco}
\end{equation}
where $S_I$ is the regulated finite on-shell string worldsheet
action whose boundary corresponds to the null-like rectangular loop $C$.
The relation $\langle {\cal W}^F(C)\rangle^2
\simeq \langle{\cal W}^A(C)\rangle$ holds
for large $N_c$ \cite{Liu:2006ug}.
To carry on the calculation based on our deformed AdS geometry,
we rotate coordinate to light-cone one as
$(t,x^1) \rightarrow (x^+,x^-)$, then the metric becomes
\begin{equation}
ds^2=\fft{1}{z^2}[-(1+f)dx^+dx^-+\fft12(1-f)[(dx^+)^2+(dx^-)^2]
+(dx^2)^2+(dx^3)^2+f^{-1}dz^2].
\end{equation}
We set the pair of quarks at $x^2=\pm\fft{L}{2}$ and choose the
worldsheet coordinates $(\tau,\sigma)$ to be
$(x^-,x^2)$.
In this setup, we can ignore the effect of $x^-$ dependence of the worldsheet
and the string is simply configured by $z(x^2)=z(\sigma)$
in the limit that $L^-$ is much larger than $L$.
Then, the string action is given by
\begin{equation}
S =\fft{R^2L^-}{\sqrt{2}\pi\alpha'z_T^2}\int_0^{L/2}{d\sigma
e^{\fft{\Phi(z)}{2}}\sqrt{1+f^{-1}{z'}^2}.}
\end{equation}
The equation of motion of $z$ can be solved by
\begin{equation}
E^{-2}=e^{-\Phi(z)}(1+f^{-1}z'^2),
\end{equation}
where $E$ is a normalized energy of motion.
Plugging the above equation into the action, we have
%In principle one can solve for $z$, but only the following integral
%will be relevant:
\begin{equation}
S=\fft{R^2L^-}{E\sqrt{2}\pi\alpha'z_T^2}
\int_0^{L/2}e^{\Phi(z)}d\sigma =
\fft{R^2L^-}{\sqrt{2}\pi\alpha'}%z_T^2} z_T^2
\int_0^{z_T}\fft{e^{\Phi(z)}dz}{\sqrt{(z_T^4-z^4)(e^{\Phi}-E^{2})}}.
\end{equation}
Then the low-energy
\footnote{The low energy approximation is
justified by observing E(L) is small at the limit $L\ll L^-$.}
effective on-shell action becomes
\begin{equation}
S\simeq \fft{R^2L^-}{\sqrt{2}\pi\alpha'}
\left(\int_0^{z_T}{\fft{e^{\fft{\Phi(z)}{2}}dz}
{\sqrt{z_T^4-z^4}}}+\fft{E^2}{2}\int_0^{z_T}
{\fft{e^{-\fft{\Phi(z)}{2}}dz}{\sqrt{z_T^4-z^4}}}\right).
\end{equation}
This action needs to be subtracted by the inertial mass of quarks,
given by two parallel flat string worldsheets along $x^-$-$z$
plane, that is,
\begin{equation}
S_0=\fft{2L^-}{2\pi\alpha'}\int_0^{z_T}{d\sigma
e^{\fft{\Phi(z)}{2}}\sqrt{G_{--}G_{zz}}}
=\fft{R^2L^-}{\sqrt{2}\pi\alpha'}\int_0^{z_T}
{\fft{e^{\fft{\Phi(z)}{2}}dz}{\sqrt{z_T^4-z^4}}}.
\end{equation}
Notice that the integral is from the AdS boundary up to the horizon.
Therefore, the net on-shell action is given by
\begin{equation}
S_I=S-S_0\simeq \fft{R^2L^-}{\sqrt{2}\pi\alpha'}\fft{E^2}{2}
\frac{F(z_T)}{z_T},
\label{SI}
\end{equation}
where
\begin{equation}\label{f1}
F(z_T)\equiv
z_T\int_0^{z_T}{\fft{e^{-\fft{\Phi(z)}{2}}dz}{\sqrt{z_T^4-z^4}}}
=z_T\fft{c^{1/2}\pi^{3/2}}{8} \left[I^2_{-\fft14}\left(\frac{c
z_T^2}{4}\right) -I^2_{\fft14}\left(\frac{c
z_T^2}{4}\right)\right], 
\end{equation}
%and $I_n(x)$ is the modified Bessel function of first kind.
$E$ is also replaced by $L$ in the following way,
%%%%%%%%%%%%%%%%%%%%%%%%%%%%%%%%%%%%%%%%%%%%%%%%%%%%%%%%%%%%%%%%%%%%%%%%%%%%%%%%%
\begin{equation}
\fft{L}{2}=\int_0^{L/2}{d\sigma}= E
z_T^2\int_0^{z_T}{\fft{dz}{\sqrt{\left(z_T^4-z^4\right)\left(e^{\Phi(z)}-E^{2}\right)}}}
\equiv E z_T f(E ; z_T). \label{f2}
\end{equation}
%where $f(E;z_T)=\int_0^{1}{\fft{dz}{\sqrt{(1-z^4)(h^2(z z_T)-E^2)}}}$.
Note that $f(E;z_T)$ is a increasing function of $E$ (regular at $E=0$),
and $f(0 ; z_T)=F(z_T)$ in Eq.~(\ref{f1}).

To determine the jet quenching parameter,
we only have to obtain the coefficient $a_1$ of linear term
in expansion $E(L)=\sum a_n L^n$.
From Eq.~(\ref{f2}) one find
\begin{equation}
a_1=\frac{\partial E}{\partial L}|_{L=0}=\frac{1}{2 z_T}
\lk f(E;z_T)+E \frac{\partial f(E;z_T)}{\partial E} \rk^{-1}|_{E=E(0)=0}=
\frac{1}{2 z_T f(0 ; z_T)}.
\end{equation}
Plugging the above expression into Eq.~(\ref{SI})
and comparing it with Eq.~(\ref{Adsco1}-\ref{Adsco}),
we obtain the jet quenching parameter,
\begin{equation}
\hat{q}
=2\fft{R^2}{2\pi\alpha'}\fft{f^{-1}(0;z_T)}{z_T^3} \\
=\sqrt{\lambda}\pi^2F^{-1}\left(1/\pi T\right) T^3.
\label{JQ1}
\end{equation}
%%%%%%%%%%%%%%%%%%%%%%%%%%%%%%%%%%%%%%%%%%%%%%%%%%%%%%%%%%%%%%%%%%%%%%%%%%%%%%%%%

For trivial $\Phi$ ($c\rightarrow 0$), we have
\begin{equation}
F^{-1}(z_T)\to \frac{\Gamma(3/4)}{\sqrt{\pi}\Gamma(5/4)},
\end{equation}
and this reproduces the result in \cite{Liu:2006ug},
\begin{eqnarray}
\hat{q}\simeq\frac{\pi^{3/2}\Gamma(3/4)}{\Gamma(5/4)}\sqrt{\lambda}T^3.
\end{eqnarray}

\begin{figure}[h]
\center{
\includegraphics{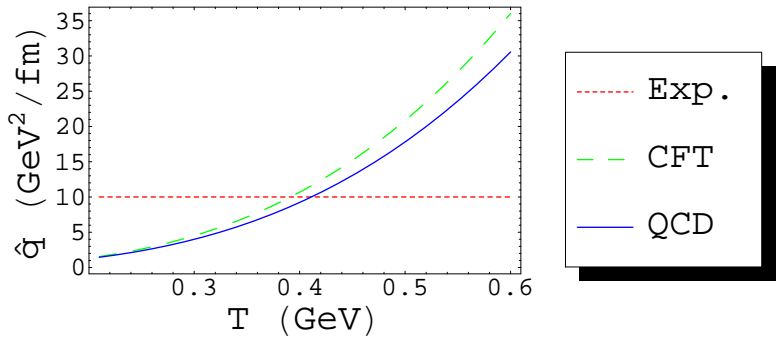}}
\caption{Jet quenching parameter. Solid curve is our result for
AdS/QCD model with $c=0.23$ GeV$^{2}$ and
$R^2/\alpha'=3.63$, and dashed one is for usual AdS/CFT model
with $\lambda=6\pi$. Straight dotted line is an evolution-time
averaged value of $\hat{q}$ from RHIC data with a range $5$-$15$
GeV$^2$/fm.} \label{jq1}
\end{figure}

Since $F(z_T)$ in Eq.~(\ref{f1}) is a decreasing function in $c$, 
the expression (\ref{JQ1}) shows that 
finite $c$ as a non-trivial dilaton effect 
enhances the jet quenching parameter for fixed $\lambda$ and temperature. 
Fig.~\ref{jq1} shows our result of jet quenching parameter. 
%, whichis smaller than the estimation in usual AdS/CFT model
Note that the figure appears to show somehow 
that our $\hat{q}$ is reduced slightly in comparison with that from AdS/CFT. 
This is due to different values for $\lambda$ we have put in:
$\left(R^2/\alpha'\right)^2=\lambda \simeq 13.2$ for ours 
from Eq.~(\ref{fitting1}), 
while $\lambda=6\pi\simeq 18.8$ for the AdS/CFT result \cite{Liu:2006ug}. 

In RHIC experiment, $\hat{q}$ decreases
temporally after the collision as temperature goes down during
expansion of QGP, and its evolution-time averaged value is
estimated to be $\bar{\hat{q}}\simeq 5-15 $GeV${}^2$/fm
\cite{jq1,Liu:2006he}, which can be reproduced from
Eq.~(\ref{JQ1}) at $T \simeq 320-470$ MeV. This range of
temperature seems consistent with what one expects for initially
equilibrated temperature $T_{0} \simeq 360$ MeV at RHIC 
\cite{jq1,Heinz:2003}.

There are suggestions in \cite{Liu:2006ug,Liu:2006he}
which account for discrepancies between $\hat{q}$ calculated from
the AdS/CFT correspondence and other non-conformal models such as
ours: the first possibility is that the number of color adjoint
degrees of freedom decreases by some factor as going from $N=4$
SYM to a generic QCD model, then $\hat{q}$ is conjectured to {\it
decrease} somewhat. 
%In agreement with this suggestion, our result
%shows that the jet quenching parameter calculated based on this
%specific AdS/QCD model in fact becomes smaller. 
Other possibilities in the context of AdS/CFT have also been 
explored\cite{jetq,Liu:2006he}: 
for example, while the $1/\lambda$ correction
will reduce the $\hat{q}$, including chemical
potential will enhance it. It may still worth exploring similar
effects in the model being concerned.

To compare our result with experimental values explicitly, we
employ the Bjorken expansion scaling \cite{BJOR1} for the
temperature evolution as $T(\tau)=T_0(\tau_0/\tau)^{1/3}$ with
$\tau_0=0.5$ fm, and an evolution-time averaged value of $\hat{q}$
defined by $\bar{\hat{q}}=\frac{4}{(L^-)^2}
\int_{\tau_0}^{\tau_0+L^-/\sqrt{2}} \hat{q}(\tau)\tau d\tau$
\cite{Liu:2006he}. Our result produces $\bar{\hat{q}}=3.5$
GeV$^2$/fm provided $T_0=360$ MeV \cite{Heinz:2003} and
$L^-/\sqrt{2}=2$ fm \cite{jq1} for hard parton travelling, while
AdS/CFT model \cite{Liu:2006he} gives $\bar{\hat{q}}=3.9$
GeV$^2$/fm with $\lambda=6 \pi$.

Regarding the experimental data from RHIC, it seems premature to
make comparison with our result, and with AdS/CFT models as well,
because of the large range of $\bar{\hat{q}}$ from the experiment.
There might be other energy loss sources beside gluon radiation,
while only the later corresponds to $\hat{q}$ calculated in this
paper. This is also suggested in \cite{Buchel:2006}.

Recently it has been argued that the string configuration used for
jet quenching calculation may not be physical
\cite{Chernicoff:2006hi,Caceres:2006ta,Argyres:2006vs,Argyres:2006yz},
thus {\sl physical} meaning of the result obtained above remains
an open question.  It may be a little early to discuss the
discrepancy between result based on AdS/CFT calculation and that
of RHIC experiment at this moment, before a better understanding
of physical configuration of a moving quark-antiquark pair inside
the QGP is provided.

\section*{Acknowledgements}
We are grateful to the TNPSS11 summer school at NTU for bringing
inspiration and collaboration on this work.  We would like to
thank J.~W.~Chen at NTU and D.~Tomino at NTNU for valuable
discussions and comments. EN would like to thank M.~Natsuume at
KEK and T.~Kunihiro at YITP for useful discussions.  We are
supported in part by the Taiwan's National Science Council and
National Center for Theoretical Science under grant
NSC94-2119-M-002-001(ST) and NSC95-2811-M-002-013(WYW). ST was
also supported in part by the Academia Sinica under grant
NSC95-2112-M-001-013.

%%%%%%%%%%%%%%%%%%%%%%%%%%%%%%%%%%%%%%%%%%%%%%%%%%%%%%%%%%%%%%%%%%%%%%%%%%%%

%%%%%%%%%%%%%%%%%%%%%%%%%%%%%%%%%%%%%%%%%%%%%%%%%%%%%%%%%%%%%%%%%%%%%%%%%%%%

\end{document}